# Switching spin valves using r.f. currents

K. Rivkin[1] and J. B. Ketterson[1,2,3]

1. Department of Physics and Astronomy, Northwestern University

Evanston IL, 60201

2. Department of Electrical and Computer Engineering, Northwestern University

Evanston IL, 60201

3. Materials Research Center, Northwestern University

Evanston IL, 60201

**Abstract**

We show that magnetization reversal in spin-injection devices can be significantly faster when using a chirped r.f. rather than d.c current pulse. Alternatively one can use a simple sinusoidal r.f. pulse or an optimized series of alternating, equal-amplitude, square pulses of varying width (a digitized approximation to a chirped r.f. pulse) to produce switching using much smaller currents than with a d.c. pulse.

There has recently been a surge of interest in new approaches to electronics (collectively called spintronics[1]) in which the operation of individual device elements is due to their interaction with a spin polarized current. Here we focus on devices that involve a spin polarized current applied perpendicular to the easy axis of a magnetic material and examine the optimal way to reverse (switch) the magnetization direction in such devices.

It is well known that the fastest way to switch the magnetization direction in conventional devices is via a uniform rotation; there are two known ways to effectively produce such switching. The first can be called *precessional switching* and the second will be referred to as *resonant switching*. In the first of these an external d.c. (flat-topped) magnetic field pulse is applied perpendicular to the easy axis. This creates a uniform Larmor precession in a plane perpendicular to the applied field. Clearly the applied field must be much larger than the effective anisotropy field. The second approach involves the application of an r.f. pulse. While it was shown experimentally by Thirion et al[2] that r.f. fields can be used together with dc fields to lower the values of dc field needed for switching, it was thought that switching with pure r.f. fields is impractical due to the excitation of the Suhl instability[3]. However we have recently shown[4] that the Suhl instability can be effectively suppressed by varying such parameters as the object size and the applied fields. In this paper we will show how the idea of switching with r.f. fields can be applied to the spin valves.

The behavior of magnetic systems under the influence of a spin-polarized current has typically been modeled using a modified Landau-Lifshitz[5] equation (here we assume the system can be approximated by a single spin which is discussed further below)

$$\frac{d\mathbf{m}}{dt} = -\gamma \mathbf{m} \times \mathbf{h} \qquad (1)$$

where $\mathbf{m}$ is the magnetization, $\gamma$ is the gyromagnetic coefficient and $\mathbf{h}$ is an effective magnetic field. We take this effective field to have the form

$$\mathbf{h} = \mathbf{h}^{true} + \frac{\mathbf{m}}{M_S} \times \left(\beta \mathbf{h}^{true} - a_J \mathbf{I}\right) \qquad (2)$$

where $\mathbf{h}^{true}$ accounts for "true" external and internal (exchange/anisotropy) fields, $\beta$ is a damping coefficient, and $M_S$ is the saturation magnetization; the interaction between the magnetization and the electrical current has here been modeled as a *spin transfer torque* [6,7,8,9,10], $\frac{\gamma a_J}{M_S} \mathbf{m} \times (\mathbf{m} \times \mathbf{I})$, where $\mathbf{I}$ is the polarization of the current and $a_J$ is an empirical factor measuring the strength of the coupling (in units of magnetic field[4] where 1000 Oe corresponds to $10^8$ A/cm$^2$).

We start by considering systems with uniaxial anisotropy; similar results can be obtained for the shape anisotropy. We introduce a coordinate system in which the z-axis lies along the easy magnetization axis.

In the resonant switching strategy[11], a small r.f. field is applied to the system. For optimum response the frequency of this field must generally be "chirped" in such a way that it tracks the system's own resonant frequency (since anisotropy is present, the resonant frequency depends on the angle the magnetization forms with the easy axis).

Although the field can be applied in the same direction as with the uniform rotation mode (i.e. perpendicular to the direction of the total magnetic field), the analytical calculations are simplest if circular polarization is assumed, with the result that the phase is shifted by $\pi/2$ with respect to the instantaneous direction of the magnetization. With this strategy the switching time is quite short and this time does not depend on the anisotropy constant. In resonant switching one needs only small effective fields.

When r.f. currents (rather than r.f. fields) are used to facilitate switching, the time dependent magnetization can be described as

$$\begin{aligned} M_x &= M_s \sin(\gamma a_J t) \sin(\varphi(t)) \\ M_y &= -M_s \sin(\gamma a_J t) \cos(\varphi(t)) \\ M_z &= M_s \cos(\gamma a_J t) \end{aligned} \qquad (3)$$

$$\varphi(t) = K \frac{\sin(\gamma a_J t)}{a_J M_s} + \gamma H_0 t \qquad (4)$$

with the components of the r.f. current given by

$$\begin{aligned} I_x &= -\sin(\gamma a_J t) \sin(\varphi(t)) \\ I_y &= \sin(\gamma a_J t) \cos(\varphi(t)) \\ I_z(t) &= -\cos(\gamma a_J t) \end{aligned} \qquad (5)$$

here K is the anisotropy coefficient and $H_0$ is the d.c. field which is applied along z-axis (note a d.c. field is not a requirement for switching and the switching time does not depend on either K or $H_0$). As can be seen, this method, while being theoretically the most effective method to produce switching, may prove complicated when it comes to practical implementation, although we offer a strategy to address this problem below.

A conventional spin valve consists of two layers, a thin one (the free layer) and a thick one (the source), separated by a nonmagnetic (injection) layer. The thick layer determines the polarization of the current, while the free layer determines whether the system is going to be in the low resistance (magnetization of the free layer parallel to the magnetization of the source layer), or in the high resistance (magnetization of the free layer anti-parallel to the magnetization of the source layer) state. Switching occurs when a sufficiently strong current is applied. Unfortunately switching requires that the magnetization of the free layer not be entirely parallel or anti-parallel to the magnetization of the free layer and therefore to the direction of polarization of the current. The cause for such misalignment is thought to arise from finite-temperature fluctuations. The switching of this configuration is explicitly non-uniform in nature.

An alternative configuration was proposed by Kent et al[12]. In this configuration there is a second "source" layer, with its magnetization oriented perpendicular to the magnetizations of the other two layers and parallel or to the direction of the current. In this case the magnetization switching is accomplished by applying a current with the polarization determined by the magnetization of this second source layer.

This latter configuration allows for resonant switching with r.f. currents. The simplest way is to use a sinusoidal r.f. current. While such a configuration is not the most effective one (e.g., the current-induced torque vanishes at one point during the rotation), we will show numerically that it is still advantageous to use an r.f. rather than a d.c. current.

Here we will use parameters similar to those used by Kiselev et al[13] in which the free layer is 2nm thick. The layer is deposited from permalloy and approximates an ellipse with principal diameters 130 by 70nm. We assume there is an exchange interaction acting between the layers. We take $A = 1.3 \cdot 10^{-6}$ erg/cm, a typical value for the exchange stiffness of permalloy; however we use $M_s = 680$ emu/cm$^3$ for the saturation magnetization, which is lower than the accepted value, but provides a better fit to the experimental results[1] (a result that may arise from the Cu layer separating the thick and thin magnetic layers). In all configurations analyzed here, an external magnetic field is assumed present that cancels the fields from the source layers. The coordinate system is chosen such that the easy axis is parallel to the z-axis with the magnetic layers lying in the y-z plane; the current direction and polarization are along x direction. In our case a current of 1 mA should correspond to $a_J \cong 144$ Oe, and the damping coefficient $\beta$ is taken as[9] 0.014.

We used two models to represent the thin and thick layers; a *multi-spin* model, with each of the magnetic layers represented by an array of individual magnetic dipoles (around 20,000 in total) and a *macro-spin* model, where the thick layers are assumed to be uniformly magnetized and static, while the free layer is described by a single spin, rotating under the influence of the field arising from the thick layer, the external current and the anisotropy field. The shape anisotropy field in the thin film is described by using the numerically calculated demagnetization tensor for an ellipse. This is somewhat different from the macrospin approach used earlier[14]; in both cases it was assumed that the free layer is an ellipse, and the value of the magnetic field due to the thick layer was determined experimentally. Comparing various results obtained by using the macro-spin model (resonant frequencies, switching speeds) with those from the multi-spin model shows that in this particular case the macro-spin model should be a good approximation.

We first attempt to switch the magnetization $+M_s\hat{z}$ to $-M_s\hat{z}$ using only a dc current. Our analysis of d.c. current-driven switching, shows that the onset is at 0.62 mA. If on the other hand we use r.f. currents with a sinusoidal time-dependence, switching is possible for much smaller currents; however one has to choose an appropriate frequency in order to obtain switching. In figure 1 we show the minimum value of $M_x / M_s$ as a function of the frequency. For our configuration the Larmor resonant frequency is 5.5 GHz, so it is apparent that switching is possible for a range of frequencies, all of them significantly smaller than the Larmor frequency. The explanation for this phenomenon is that in objects with shape anisotropy the resonant frequency depends on the angle between the magnetization vector and the anisotropy axis; for large precession angles this frequency can differ significantly from the small-angle Larmor frequency. Assuming the magnetization is initially aligned along the easy axis, the resonant frequency will in general decrease as the magnetization tips away from the easy axis; therefore the optimal switching frequency will clearly be lower than the small-angle resonant frequency.

Calculating the lowest frequency that allows switching as a function of the r.f. current amplitude, we obtain a nearly linear behavior of the form:

$$\omega = 3.74 - 4.015 \cdot I \, . \tag{6}$$

For currents smaller than I = 0.21 mA switching no longer occurs; on the other hand there is no lower limit (in the absence of damping) with the chirped-pulse strategy. However this current is still almost 3 times lower than the d.c. current needed to switch this system.

In figure 2 we show the time required for switching $M_z$ with d.c. vs. r.f. currents. In both cases, if the currents are not turned off the system will continue to oscillate. A d.c. current allows for faster switching (albeit with much higher currents), but it also stays in the region where $M_z$ is negative for only very short times, which would require very precise pulse lengths. On the other hand, switching via r.f. currents results in the system having negative values of $M_z$ approximately 15 times longer than in the case when d.c. currents are used.

As with the magnetically driven case, one anticipates that chirping the r.f. current will yield more efficient switching. However, rather than using a constant-amplitude sinusoidal r.f. pulse with a chirped phase function, we have found that a series of square (flat-topped) pulses of equal magnitude, alternating sign, and varying width produces very stable switching; such a waveform might be synthesized digitally in practice. The associated pulse widths are chosen so as to maximizes the decrease in $M_z$: From Eqs. (1) and (2) it follows that the rate of change of $M_z$ due to the applied current is:

$$\frac{dM_z}{dt} = \gamma a_J(t) \frac{M_x M_z}{M_s} \qquad (7)$$

hence if we wish to minimize this term with the chosen waveform, the sign of $a_J$ should be opposite to that of $M_x M_z$.

In figure 3 we present the resulting pulse sequence and the behavior of $M_z / M_s$ vs. time for the case where the current amplitude is 0.17 mA. Figure 4 shows a plot of the current reversal times as a function of the current amplitude, starting from the lowest value for which switching is still possible – 0.089 mA. The current is switched off when the switching is completed.

In conclusion we have shown that the fastest available methods for switching traditional magnetic devices, (precessional and chirp-resonant), can also be used for spin valves. While implementing such methods is demanding, we have shown that some simpler methods, utilizing a sinusoidal r.f. current pulse or an optimized sequence of sequence of alternating square current pulses, can lower the required switching currents.

**Acknowledgments.**

This work was supported by the National Science Foundation under grants ESC-02-24210 and DMR 0244711.

# References

[1] J. Slonczewski Journal of Magnetism and Magnetic Materials **159**, L1 (1996).

[2] C. Thirion, W. Wernsdorfer, D. Mailly "Switching of magnetization by non-linear resonance studied in single nanoparticles", Nature Mat. **2**, 524-527 (2003).

[3] P.W. Anderson and H.Suhl "Instability in the motion of ferromagnets at high microwave power levels", Physical Review **100**, 1788 (1955).

[4] K.Rivkin, V.Chandrasekhar and J.B.Ketterson "Controlling the Suhl instability: a numerical study", submitted to Physical Review Letters.

[5] L. D. Landau and E. M. Lifshitz, Phys. Z. Soviet Union **8**, 153 (1935).

[6] A. Brataas, Y.V. Nazarov and G. E. Bauer, Physical Review Letters **84**, 2481 (2000).

[7] S. Zhang, P. M. Levy and A. Fert, Physical Review Letters **88**, 236601 (2002).

[8] J. E. Wegrove et al Europhysics Letters **45**, 626 (1999).

[9] Z. Li and S. Zhang Physical Review B **68**, 024404 (2003).

[10] J. Z. Sun Physical Review B **62**, 570 (2000).

[11] K. Rivkin and J. B. Ketterson "Magnetization reversal in the anisotropy-dominated regime using time-dependent magnetic fields", submitted to Science.

[12] A. D. Kent, B. Ozyilmaz and E. del Barco, Applied Physics Letters **84**, 3897 (2004).

[13] S. I. Kiselev, J. C. Sankey, I. N. Krivorotov, N. C. Emley, A. G. F. Garcia, R. A. Buhrman and D. C. Ralph "Spin Transfer Excitations for permalloy nanopillars for large applied currents", http://www.arxiv.org/list/cond-mat/0504402.pdf

[14] S. I. Kiselev, J. C. Sankey, I. N. Krivorotov, N. C. Emley, R.J. Schoelkopf, R.A. Buhrman and D. C. Ralph Nature **425**, 380 (2003).

**Figure Captions**

**Figure 1. Minimal value of $M_Z/M_S$ as a function of the frequency of r.f. current.**

**Figure 2. $M_Z/M_S$ as a function of time for switching with r.f. and d.c. currents.**

**Figure 3. The magnetization $M_Z/M_S$ as a function of time for switching with an optimized sequence of square pulses with amplitude $I_0 = 0.17$ mA.**

**Figure 4. Current reversal times for a sequence of square pulses as a function of current amplitude.**

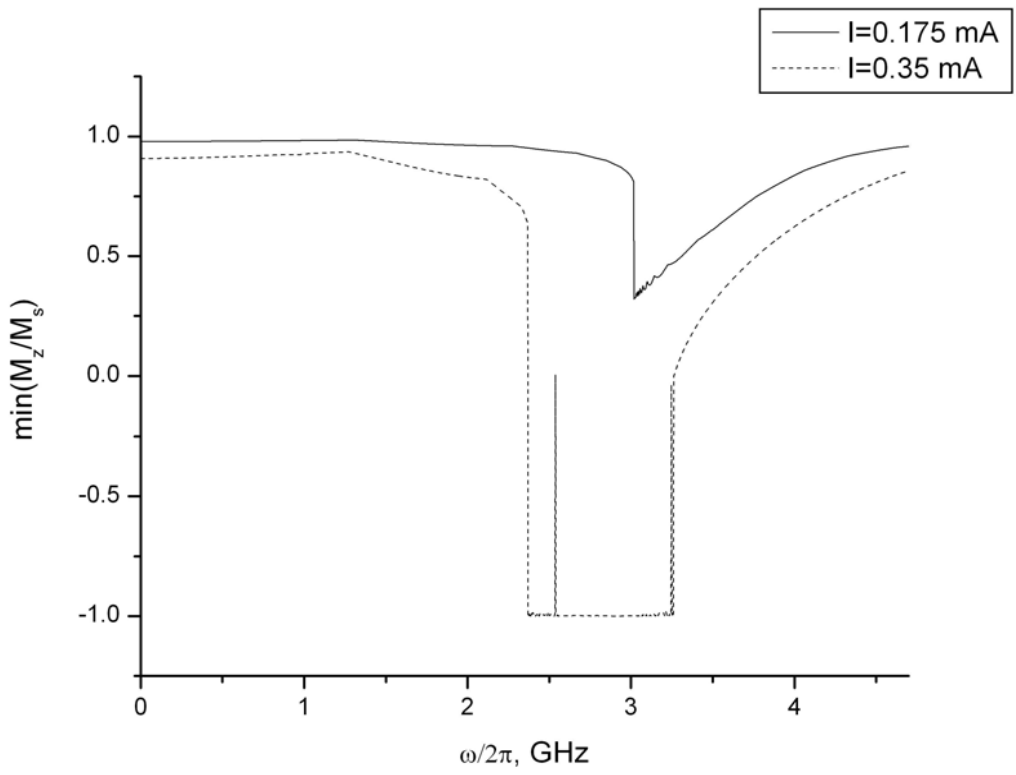

**Figure 1.** Minimal value of $M_z / M_s$ as a function of the frequency of r.f. current.

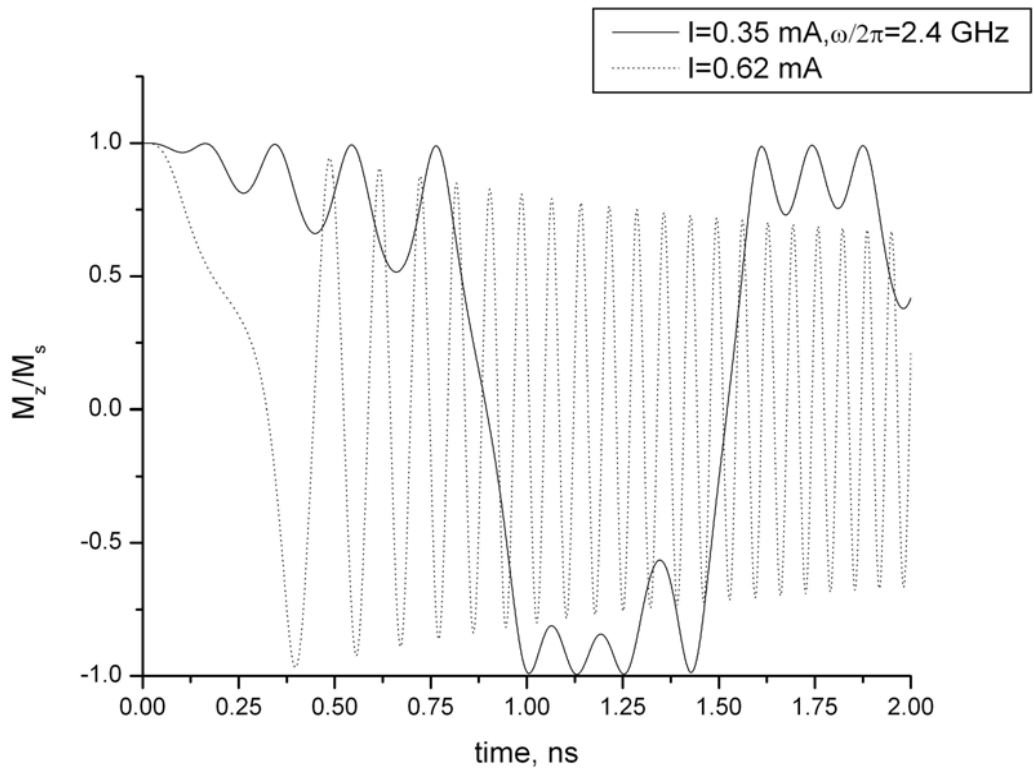

**Figure 2.** $M_z / M_s$ **as a function of time for switching with r.f. and d.c. currents.**

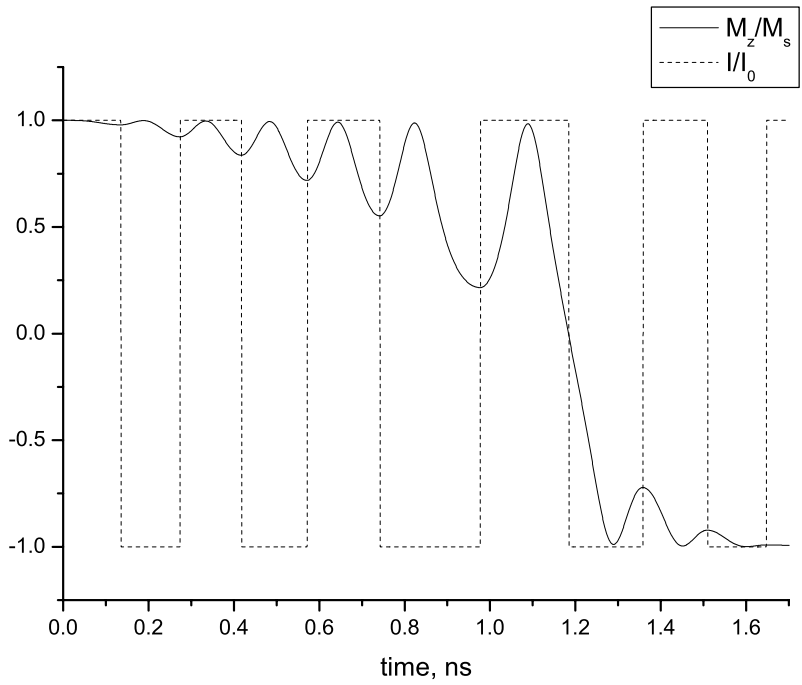

**Figure 3. The magnetization $M_z / M_s$ as a function of time for switching with an optimized sequence of square pulses with amplitude $I_0 = 0.17$ mA.**

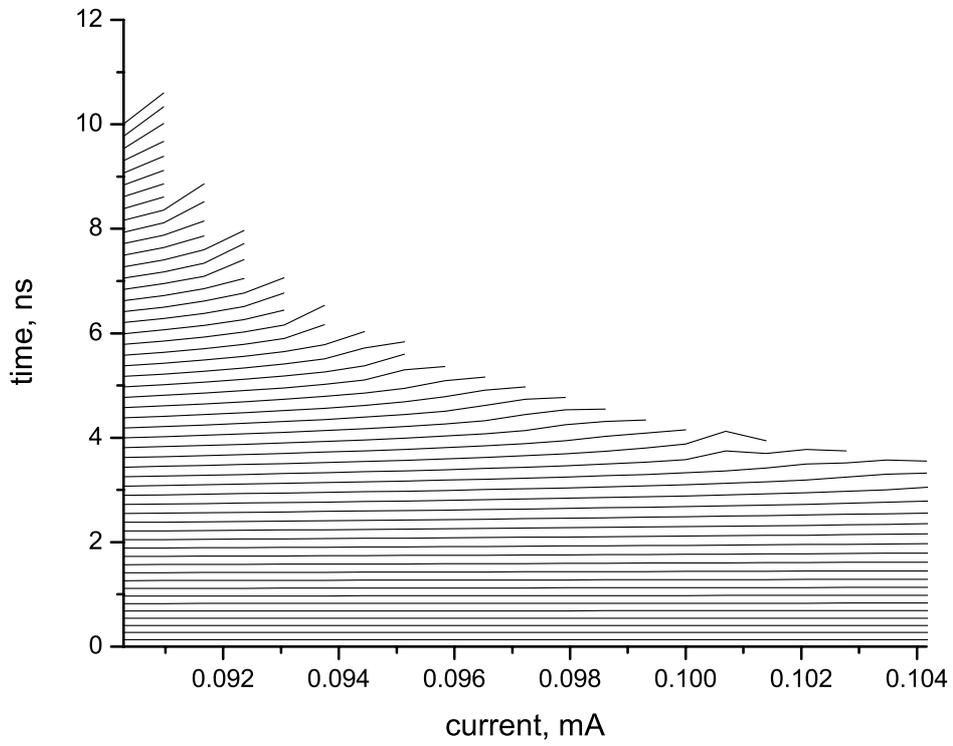

**Figure 4. Current reversal times for a sequence of square pulses as a function of current amplitude.**